\documentclass[letter]{aa} 

\usepackage{lipsum}
\usepackage{natbib}
\usepackage{amsmath}
\usepackage{xspace}
\usepackage[utf8]{inputenc}
\usepackage{txfonts}
\usepackage{multirow}

\usepackage{xspace} 

\newcommand{\msun}{\ensuremath{\mathrm{\,M_\odot}}} 
\newcommand{\rsun}{\ensuremath{\mathrm{\,R_\odot}}} 
\newcommand{\kms}{\ensuremath{\mathrm{\,km\,s^{-1}}}} 

\newcommand{\mco}{\ensuremath{M_{\mathrm{CO}}}} 
\newcommand{\mdot}{\ensuremath{\dot{M}}} 

\newcommand{\mesa}{{\textsc{mesa}}\xspace} 

\definecolor{myblue}{cmyk}{1,1,0,0} 
\usepackage[colorlinks,citecolor=blue,urlcolor=blue]{hyperref}
\graphicspath{{./}{figures/}}
\usepackage{xcolor}
\definecolor{ochre}{rgb}{0.8, 0.47, 0.13}
\definecolor{mygreen}{rgb}{0.19,0.55,0.11}

\makeatletter
\@fpsep\textheight
\makeatother
\makeatletter
\renewcommand*\aa@pageof{, page \thepage{} of \pageref*{LastPage}}
\makeatother
\begin{document}

\title{Explodability fluctuations of massive stellar cores enable asymmetric compact object mergers such as GW190814}

   \author{John Antoniadis\inst{1,2,3}
   \and  David R. Aguilera-Dena\inst{1,3,2}
   \and Alejandro Vigna-G\'omez\inst{4,5}
   \and Michael Kramer\inst{2,6}
   \and Norbert Langer\inst{3,2}
   \and Bernhard M\"uller\inst{7}
   \and Thomas M. Tauris\inst{8}
   \and Chen Wang\inst{3}
   \and Xiao-Tian Xu\inst{3}
    }

   \institute{Institute of Astrophysics, Foundation for Research and Technology--Hellas, N. Plastira 100, Voutes, GR-71003, Greece
    \and
    Max-Planck-Institut f\"ur Radioastronomie, Auf dem H\"ugel 69, 53121 Bonn, Germany
    \and
    Argelander-Institut f\"ur Astronomie, Universit\"at Bonn, Auf dem H\"ugel 71, 53121 Bonn, Germany 
    \and
    DARK, Niels Bohr Institute, University of Copenhagen, Jagtvej 128, 2200, Copenhagen, Denmark
    \and
    Niels Bohr International Academy, The Niels Bohr Institute, Blegdamsvej 17, 2100 Copenhagen, Denmark
    \and
    Jodrell Bank Centre for Astrophysics, University of Manchester, Manchester M13 9PL, UK
    \and 
    School of Physics and Astronomy, Monash University, VIC 3800, Australia
    \and
    Department of Physics and Astronomy. Aarhus University. Ny Munkegade 120. DK-8000 Aarhus, Denmark
    }
   

  \abstract{The first three observing runs with  Advanced LIGO and Virgo  have resulted in the detection of binary black hole  (BBH) 
  mergers with highly unequal mass components, which are 
  difficult to reconcile with standard formation paradigms. 
 The most representative of these is GW190814, a highly asymmetric 
  merger between a 23\msun~black hole (BH) and a 2.6\msun~compact object.   
  Here, we explore recent results, suggesting that a sizable fraction of stars with pre-collapse  carbon-oxygen core masses above $10\msun$, and extending up to at least $30\msun$, 
  may produce objects inside 
  the so-called lower mass gap that bridges the division between massive pulsars and BHs in Galactic X-ray binaries. We demonstrate that such an explosion landscape would naturally cause a fraction of massive binaries to produce GW190814-like systems instead of symmetric-mass BBHs. We present examples of specific evolutionary channels leading to the formation of GW190814 and GW200210, a $24+2.8\msun$ merger discovered during the O3b observing run. We estimate  the merger-rate density of these events in our scenario to be $\mathcal{O}$(5\%) of the total BBH merger rate. Finally, we  discuss the broader implications of this formation channel for compact object populations,  and  its possible relevance to less asymmetric merger events such as GW200105 and GW200115.}

\titlerunning{Formation of GW190814}
\authorrunning{J. Antoniadis et al.}
\maketitle

\section{Introduction}
Following the detection of GW150914 --- the first binary black hole (BBH) merger 
observed in gravitational waves \citep[GWs;][]{Abbott:2016blz} --- the three  
Advanced LIGO--Virgo Collaboration (henceforth LVC)  observing runs  
\citep[O1--3;][]{TheLIGOScientific:2014jea,TheVirgo:2014hva} 
have uncovered a diverse collection of GW signals originating in collisions 
between black holes (BHs) and neutron stars 
\citep[NSs;][]{Abbott:2016nmj,TheLIGOScientific:2016pea,TheLIGOScientific:2017qsa,Abbott:2021abc}.

GW190814, discovered during O3 \citep[][]{Abbott:2020khf}, is an atypical event whose origin is  particularly challenging to explain.  
Assuming  no  magnification due to gravitational lensing took place along 
the line-of-sight \citep{broadhurst2020}, the  signal corresponds to a merger between a 
$(23\pm 1)\msun$ and a $(2.6\pm0.1)\msun$ object. 
While the massive primary is undoubtedly a BH, the nature of the 
secondary is uncertain and not constrained by the GW signal itself: 
its  mass securely places it inside the observed `lower mass gap' (LMG) 
that separates the most massive  NSs in binary pulsar systems  
\citep[$\lesssim 2.1\msun$;][]{Ozel:2012ax,Antoniadis:2013pzd,antoniadis2016,Fonseca:2021wxt}  
from the lightest BHs in X-ray binaries \citep[XRBs; $\gtrsim 
5\msun$;][]{Ozel:2010su,Farr:2010tu}.
If this object is a NS of $\sim2.6\msun$, its high mass,
combined with previous constraints on NS properties \citep{Ozel:2016oaf,GBM:2017lvd}
would place very stringent constraints on the behaviour of matter at 
extreme densities  \citep[for detailed discussions see e.g.][]{Tsokaros:2020hli,Nathanail:2021tay}.
The possibility of a light BH instead is equally intriguing and forces us to 
rethink how such objects might form.
More recently, the LVC announced the GWTC-3 catalogue which contains additional asymmetric systems  \citep{Abbott:2021abc}. 
 One event that stands out due to its similarity to GW190814 is GW200210, a $\sim$24 + $2.8\msun$ compact-object merger. Even though this was a relatively low-significance detection  (false-alarm-rate > 1\,yr$^{-1}$), its discovery  suggests that such asymmetric systems may be relatively common. 

The combination of an extreme mass ratio  and  odd secondary mass in these systems poses an additional challenge for standard  binary and dynamical formation channels.  
While certain scenarios are able to produce GW190814-like systems,
they are either not common enough to match the empirically inferred merger-rate density  of  
$ 1 - 23$\,Gpc$^{-3}$\,yr$^{-1}$, or 
incompatible with the  birth rates of other astrophysical 
populations \citep[][see Section~\ref{sec:2}]{Kruckow:2018slo,Zevin:2020gma,Mandel:2020cig}.  

Here, we revisit isolated binary evolution channels in light of 
recent results suggesting that the 
landscape of core-collapse supernovae (CCSNe) may be substantially more 
complex than those predicted by standard binary population synthesis (BPS) prescriptions 
\citep[][]{Kruckow:2018slo,Zevin:2020gma,Mandel:2020cig}.
The text is organised as follows: Section~\ref{sec:2} reviews the properties of GW190814, 
as well as the proposed binary formation channels and their bottlenecks. 
In Section~\ref{sec:3}, we investigate the potential link between GW190814 and 
 CCSNe originating from  massive stars that would normally be expected to form high-mass BHs. 
In Section~\ref{sec:4}, we present examples of specific formation channels and explore 
their implications for the merger-rate density of GW190814-like systems. We conclude with a summary  in Section~\ref{sec:5}.

\section{Implications of GW190814-like mergers for binary evolution}\label{sec:2}
With a mass ratio of $q\simeq 0.1$, GW190814 and GW200210 are the most asymmetric compact
binary mergers discovered to date.
The observed GW190814 signal provided stringent constraints on the 
effective spin parameter, $|\chi_{\rm eff}| \lesssim 0.063$, the magnitude of spin 
precession, ${\chi_{\rm p}}\leq 0.08$, and the  spin magnitude of the more massive component, 
$\chi_1 \leq 0.06$ \citep[see][for details]{Abbott:2020khf}. 
Taken together, these estimates suggest that the primary was slowly rotating 
with a spin nearly aligned with the orbital angular momentum. These spin properties deviate from the general trend that has been found for less asymmetric systems, in which there seems to be a negative correlation between $q$ and $\chi_{\rm eff}$, with more asymmetric systems having higher effective spins  \citep{Callister:2021abc}.

Assembly via dynamical channels, for instance 
in a non-segregated cluster 
\citep{Clausen:2014ksa,Fragione:2020wac,Rastello:2020sru}, hierarchical mergers 
in multiple systems \citep{Liu:2021abc,Lu:2020gfh}, and binary mergers near galactic nuclei 
\citep{McKernan:2020lgr,Yang:2020xyi} may provide a viable formation pathway. 
However, these  generally struggle to produce such highly asymmetric 
binaries at sufficient rates \citep[see][]{Abbott:2020khf}. 

Isolated binary evolution channels producing GW190814-like mergers have been 
extensively investigated by \cite{Zevin:2020gma}, who 
explored a broad range of  physical assumptions  
using  BPS tools \citep[see also][]{Mandel:2020cig,Patton:2021svb,Shao:2021dbg}. 
Such systems might be produced via 
two broad binary formation channels: one in which the massive 
component forms first (henceforth Channel~A) 
and another in which it forms last (Channel~B).
For Channel~A, \cite{Zevin:2020gma} identify the representative progenitor 
system to be a binary consisting of a $\sim$30\msun~ zero-age main-sequence (ZAMS) 
star orbiting a 21\msun~star ($q\simeq 0.65$) in a wide,  
 $\sim$5000-day, $e\simeq 0.3$ orbit. The primary loses mass via stable, 
 non-conservative Case\,B mass transfer \citep{ywl10,tv22} 
that results in a moderate shrinkage of the orbit. The binary then receives a 
strong natal kick during the formation of the massive BH 
(post-collapse eccentricity of $e=0.98$). 
This is followed by a second episode of stable mass transfer onto the BH,  
and a  supernova (SN) that forms the least-massive component and 
forces the binary into a more compact configuration via a second natal kick  ($e=0.99$). Overall, the combination of two explosions, both leading to highly eccentric post-SN configurations, is statistically quite rare.  

In Channel~B, the typical progenitor is found to be a 
more compact binary consisting of a $\sim$45\,M$_{\odot}$ 
and a $\sim$23\,M$_{\odot}$ ZAMS star in a 2-day orbit. 
The system experiences two mass-transfer episodes 
that result in a mass inversion prior to the 
formation of the first compact object. 
The binary is kicked into a wider orbit during the SN (post-SN of $e=0.97$) 
and subsequently experiences a common envelope (CE) episode before the 
formation of the more-massive BH. A similar channel has been identified by 
\cite{Mandel:2020cig}, who conclude that this path is only possible 
when Hertzsprung-gap donors are optimistically assumed to survive CE evolution  \citep{Klencki:2020kxd}. 

\cite{Zevin:2020gma} find that at low metallicities, 
both  pathways operate at nearly equal rates, 
while at higher  metallicities, Channel~A dominates.
Their estimates depend  on  
the CE efficiency, and the relation between ZAMS and compact-object masses. 
Matching the empirical GW190814-like merger-rate density  
is extremely challenging, regardless of the underlying assumptions.  

In the context of the \cite{Fryer:2011cx} SN prescription employed in 
\cite{Zevin:2020gma},  the mass of the secondary 
can only be explained via a delayed CCSN mechanism in which instabilities 
grow over timescales $\gtrsim 200$\,ms, thereby allowing for more matter to be 
accreted onto the proto-NS compared to explosions in which such instabilities develop rapidly ($\lesssim 50$\,ms). 
While this  naturally populates the LMG in all astrophysical populations, 
the initial binary configurations and required interactions that lead to the formation of 
highly asymmetric pairs are still highly improbable. 
The main reason is that LMG objects are only relevant to progenitors 
within a narrow-mass range, with pre-collapse carbon-oxygen (CO) 
core masses (\mco) of approximately $\sim$4\msun~ \citep[see ][and Figure~\ref{fig:1}]{Fryer:2011cx}. 
Such stars have a very small a priori probability of surviving the SNe  and 
CE episodes required to form asymmetric compact binaries \citep[e.g.][]{ktl+16}. 
In the following section, we explore how a  physically motivated 
initial-to-remnant mass relation might help overcome this bottleneck.

\section{The birth mass distribution of compact objects}\label{sec:3}
As discussed in Section~\ref{sec:2},  
rate estimates for GW190814-like mergers in BPS studies 
are severely limited by the  range of stars  
that can produce compact objects inside the LMG. 
Consequently, a modified explosion landscape that 
increases the diversity of progenitors for the low-mass
component compared to standard prescriptions \citep[e.g.][]{Hurley:2000pk,Fryer:2011cx}  
may significantly alter  this picture.  

Indeed, several recent studies suggest that 
the initial-to-remnant mass relations traditionally employed in
BPS simulations might not be representative 
of the true explosion landscape \citep[][]{OConnor:2010moj,Ugliano:2012fvp,
Sukhbold:2013yca,Pejcha:2014wda,Ertl:2015rga,Muller:2016ujh,Ebinger:2019abc,Woosley:2020dsf,
Ertl:2019zks,Schneider:2020vvh,Patton:2020tiy,Mandel:2020cig,Shao:2021dbg,Zapartas:2021dfu, 
Patton:2021svb,Laplace:2021abc,aguilera-dena2021}. 
Whether a star forms a NS or a BH within the neutrino-driven explosion paradigm 
is largely determined by the pre-SN density 
\citep[or equivalently, the 'compactness';][]{OConnor:2010moj} of the 
stellar core. In turn, the latter appears to depend sensitively on how 
the late stages of nuclear burning proceed, both inside the core and in the envelope \citep[e.g.][]{Brown:2001ua,yta+21}. 
\cite{OConnor:2010moj}, \cite{Ugliano:2012fvp}, and \cite{Sukhbold:2013yca}, employing simple 
explodability criteria, first showed  
that the complex processes taking place 
during these late evolutionary stages can lead to rapid alternations between successful explosions and 
implosions within certain mass ranges. More recently, it has been shown that  `islands of  explodability'   
might be even more pronounced in stars that are stripped of their hydrogen envelopes, for example due to binary 
interactions \citep{Ertl:2019zks,Schneider:2020vvh,Mandel:2020cig,Laplace:2021abc}. 

\begin{figure*}
\resizebox{\hsize}{!}{\includegraphics{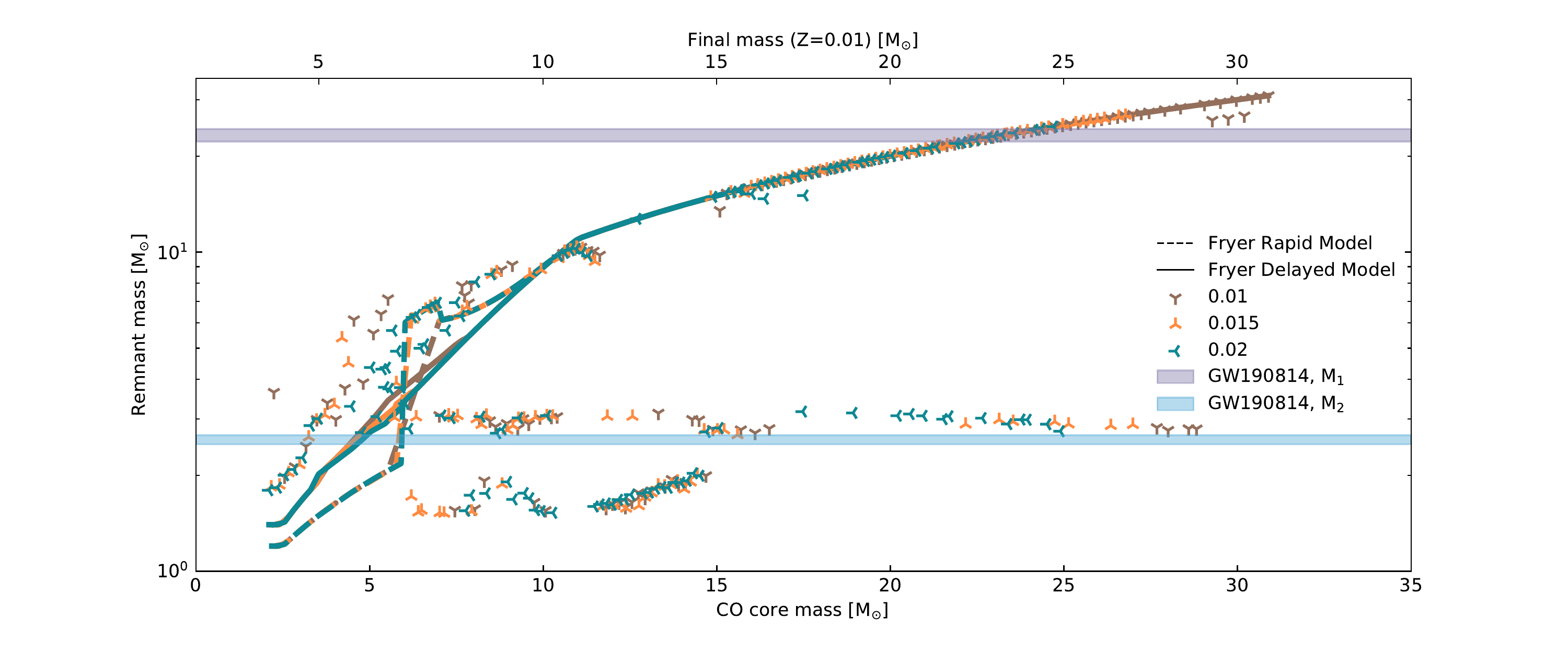}}
\caption{Relation between remnant, CO-core, and pre-collapse masses
derived from the helium-star models of \cite{aguilera-dena2021} 
using the semi-analytic explosion models of \cite{Muller:2016ujh} as 
updated in \cite{Mandel:2020abc}. The scatter plot illustrates the predictions of the aforementioned 
models for different initial metallicities ($Z=0.01,0.015,0.02$ shown in brown, orange, and green, respectively). 
Lines indicate the predictions of 
\cite{Fryer:2011cx} and their variation with metallicity. 
Horizontal strips show the inferred component masses of GW190814. 
As can be seen, detailed models predict a much broader rage of 
possible progenitors for the low-mass component compared to standard 
BPS prescriptions. Notably, two CO~cores of equal mass ($\sim24\msun$) 
may reproduce both GW190814 components. }\label{fig:1}
\end{figure*}

Recently, \cite{aguilera-dena2021} \citep[henceforth AD21; see also][]{Aguilera-Dena:2021abc}  demonstrated, for the first time,  
that  such pre-SN compactness variations extend  
to pre-collapse  CO-core masses significantly above 10\msun, suggesting that 
successful CCSN explosions might occur even for the most massive stars. 
Figure~\ref{fig:1}  illustrates  the correlations between remnant,  
CO-core, and  pre-SN  inferred by AD21. 
These relations are based on 1D evolutionary models of helium stars, 
calculated at different metallicities  using the \mesa code \citep{Paxton:2010ji,Paxton:2019abc}. 
Remnant masses were inferred using a revised version of the \cite{Muller:2016ujh}  
 parametric explosion models \citep[see][]{Mandel:2020abc}, which considers  
 the effect of SN fallback. 
 These semi-analytic prescriptions are motivated by 3D neutrino-driven simulations and 
 rely on the pre-SN stellar structure to predict 
 explosion properties and remnant masses 
 (for a more detailed discussion on the model and the input parameters 
 used to infer the quantities in Figure~\ref{fig:1}, see AD21). 
As can be seen, the predictions of AD21 differ substantially 
from analytic BPS prescriptions \citep[e.g.][]{Hurley:2000pk,Fryer:2011cx}. 
Contrary to the latter, both NSs and LMG objects are obtained for 
stars with a final $\mco$ extending up to $\sim30\msun$. 
The mass range for which explosions occur appears to depend on  metallicity; for example,  for 
$Z=0.02$, one obtains explosions for $\mco\leq25\msun$, 
while for $Z=0.01$ one finds explosions extending to $\mco\leq30\msun$. 
Most notably, our models suggest the presence of an 
extended and well-defined explodability island in the \mco=10\ldots15\msun~range \citep[first identified by][]{Schneider:2020vvh},
where CCSNe (leading to NS formation) are significantly favoured over implosions (leaving behind BHs). 
For $\mco\ge15\msun$, while the most likely outcome is an implosion,
we predict that a sizable fraction of stars 
(10\% and 20\% of our models at $Z=0.01$ and 0.02, respectively) 
produce a fallback CCSN that leads to the formation of a LMG object, instead of a massive BH. 
These remnants have masses between 2.6 
and 3.1\msun, i.e. quite similar to the low-mass component in GW190814.
Here, for simplicity, 
we assume all LMG objects to be BHs, given the current constraints on 
the equation-of-state. This choice affects our results only moderately, as
one would expect somewhat higher kick velocities (up to $\sim~1200\kms$) if the remnant is a NS.

The aforementioned properties make  high-mass stars particularly 
appealing as GW190814 progenitors for two reasons.
Firstly, due to underlying dependence on metallicity 
and the shape of the initial mass function \citep[IMF;][]{Kroupa:2000iv}, 
they are expected to be rare, and hence not particularly 
relevant to the LMG in Galactic populations. 
Secondly, despite their rarity they are very similar 
to the stars expected to form the 
more massive component. 
Such progenitors are extremely similar to those expected to form more 
`canonical' symmetric binary BHs  \citep{LIGOScientific:2020kqk}. 
Hence, they  have increased 
chances of  producing a successful merger 
\citep[e.g.][]{Belczynski:2016obo,Kruckow:2018slo,Vigna-Gomez:2018dza} 
 compared to the systems identified by  
 \cite{Zevin:2020gma} and \cite{Mandel:2020cig}. 
In the following section, we give specific examples of symmetric binaries 
that may lead to the formation of GW190814 based on the explosion landscape discussed here. 

\section{Binary formation channels for GW190814}\label{sec:4}
In this section, we demonstrate the impact of a non-monotonic relation between  pre-SN and remnant masses on the formation of  highly asymmetric mergers, using two specific evolutionary examples. While both paradigms are subject to several physical uncertainties (e.g. related to the modelling of Case~A mass transfer in BPS codes, or the outcome of CE events), they serve to demonstrate that GW190814-like systems and symmetric BBHs ($q\simeq 1$) may form in a very similar way.

\begin{figure}
\resizebox{1.05\hsize}{!}{\includegraphics[scale=1.0]{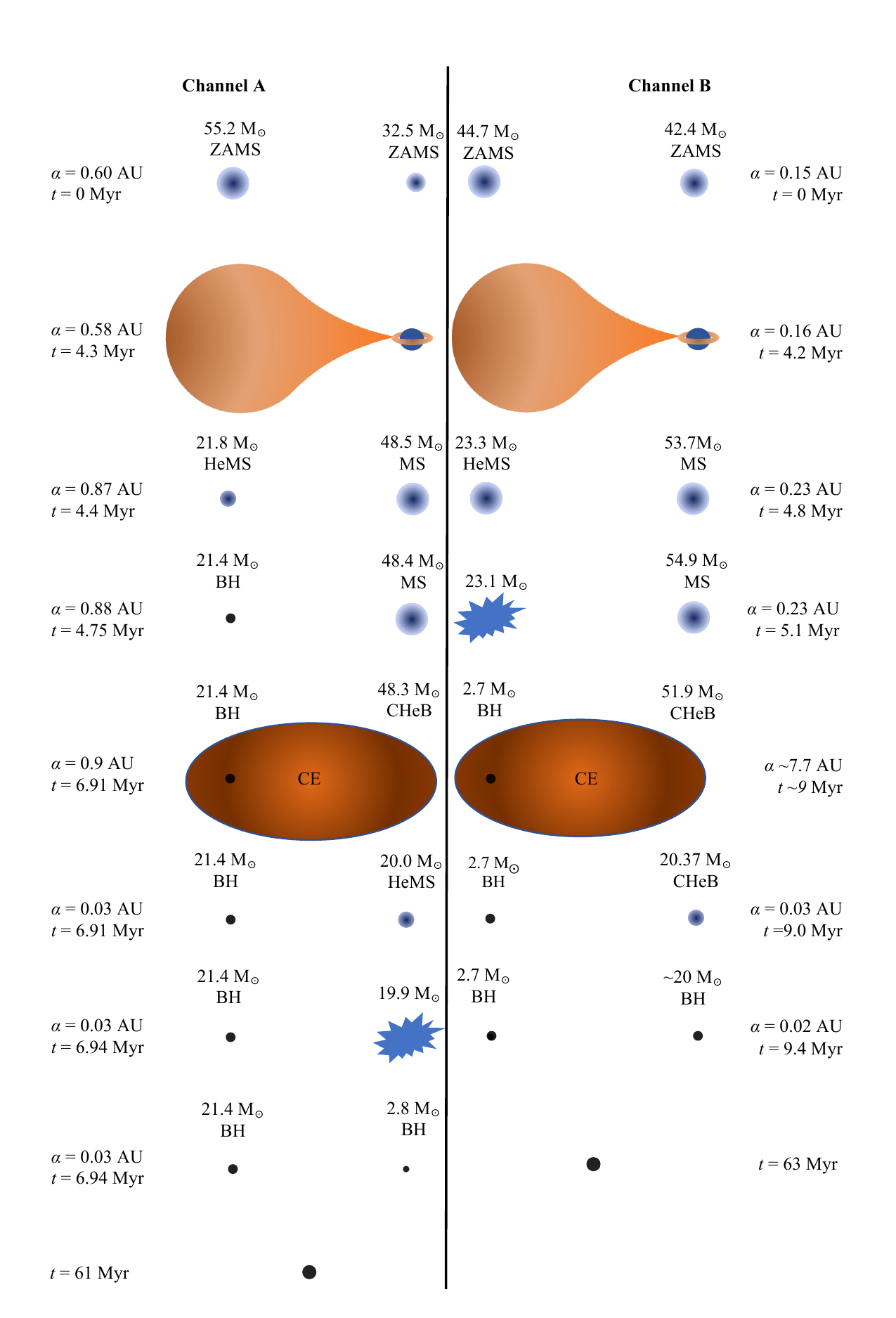}}
\caption{Examples of binary evolution leading to the formation of 
asymmetric-mass GW190814-like mergers.} \label{fig:2}
 \end{figure}

Figure~\ref{fig:2} summarises the evolution of these example systems,  
assuming the AD21 explosion landscape discussed in 
Section~\ref{sec:4} and  Fig.~\ref{fig:1}. 
The left-hand side shows a Channel\,A-type progenitor
(see also Fig.~\ref{fig:a1}) that 
was calculated  using \texttt{COMPAS} 
\citep{COMPAS:2021methodsPaper}. By employing this rapid BPS code, we were able to rapidly identify this particular example among several models, using  methods and assumptions similar to those adopted by previous population-synthesis studies \citep[e.g.][]{Zevin:2020gma}. 

The initial system has a very low metallicity (1\% solar) and  
consists of a 55.2+32.5\msun~ZAMS pair, separated by 0.6\,AU. 
The system  undergoes stable Case\,A mass transfer,   
which is followed by the formation of a 21.4\msun~BH from the initially more massive star. 
The secondary then expands during core helium burning (CHeB). 
This leads to a CE episode that expels the envelope and hardens the binary.  
The resulting system evolves without experiencing 
further mass transfer, until the secondary collapses. 
Based on the default \cite{Fryer:2011cx}   remnant-mass prescription employed in \texttt{COMPAS}, 
the resulting system would be a `canonical' 21.4+19.5\msun~binary BH system. 
Such symmetric  BBH mergers  are expected to be the most common and 
may originate from a broad range of initial configurations 
\citep[e.g.][]{Belczynski:2016obo,Stevenson:2017tfq}, in addition to the one discussed here. 
However, as shown in Section~\ref{sec:3}, the secondary might instead produce a LMG object in a fallback CCSN at the 10$-$20\% probability level, thereby creating a GW190814-like system. 

Even though this demonstrates that  symmetric- and asymmetric-mass 
mergers could have nearly identical formation histories, one key 
difference that can set them apart is the magnitude of the natal kick 
received during the CCSN. 
Our models predict a wide range of kick velocities for fallback 
CCSNe extending from 0 to $\sim$1200\kms. 
This diversity is an intentional feature of our semi-analytic SN prescription  that is physically motivated by the results of detailed 3D simulations \citep[see ][ and AD21 for details]{Mandel:2020abc,Chan:2020}. 
For stellar structures that yield low initial explosion energies, one expects 
the asymmetric inner ejecta to fall back completely. A low-energy explosion then results from a weak sound pulse that is launched when the initial ejecta become 
subsonic. The transition to the subsonic regime means that the transport of energy by the sound pulse  decouples from the transport of matter (with the initial ejecta being accreted onto the black hole). The sound pulse quickly 
becomes spherical resulting in symmetric explosions with negligible natal kicks.
Alternatively, if the initial ejecta has sufficiently high energy, it  
escapes and the momentum asymmetry between the ejecta and the remnant is  
largely  preserved, resulting in a strong kick. 

Evidently, symmetric explosions would not significantly affect the binary 
dynamics, as they may only modify the eccentricity by an amount proportional 
to the mass lost in the explosion ($e = \delta M/M \simeq 0.7$ for the example 
in Fig.~\ref{fig:2}, where $\delta M$ is the ejected mass, and $e$ and $M$ are 
the post-SN eccentricity and total mass of the binary, respectively). As long as 
$\delta M$ is smaller than the total mass, the system  remains bound and  merges on a similar timescale compared to circular systems with comparable separations. 
On the other hand, CCSNe that produce large kicks may completely alter the binary dynamics \citep[see][and references therein]{Tauris:2017omb}. Assuming  
the largest kick velocity obtained for our models ($w=1200\kms$) and a random 
kick orientation, 
for the example in Fig.~\ref{fig:2}, one finds a $34\%$ probability of surviving 
and merging within a Hubble time (see also Fig~\,\ref{fig:a3}). This estimate is  likely extremely conservative, as nearly all fallback CCSNe in our calculations produce smaller kicks. 

To summarise, given the similarities to channels that form symmetric-mass BH pairs, we believe Channel~A to be the dominant contributor of GW190814-like systems, especially at low metallicities. 
A 0th-order semi-empirical estimate for the expected  rates under our scenario 
can be obtained from the observed BBH merger rates. 
\cite{LIGOScientific:2021psn} estimate the total merger rate for BBHs to be 
 $R_{\rm BBH} = 16-130$\,Gpc$^{-3}$\,yr$^{-1}$ (90\% credible interval).
If the majority of GW190814-like systems form in a similar way to symmetric BBHs, then their expected rate would be the following:

\begin{equation}\label{eq:1}
    R_{\rm GW190814} \simeq R_{\rm BBH} \times F_{\rm fallback} \times P_{\rm survive},
\end{equation}
 where $F_{\rm fallback}$ is the IMF-weighted fraction of stars 
that produce fallback SNe instead of massive BHs, and  $P_{\rm survive}$ is the population-averaged  probability for such systems to 
survive the natal kick and produce a successful merger. Therefore, 
if $F_{\rm fallback}\simeq \mathcal{O}(10\%)$, as suggested by our models,
and $P_{\rm survive} \simeq \mathcal{O}(50\%)$ similarly to the 
(conservative) example above, one obtains
$R_{q\neq 1}\simeq 0.75-6.5$\,Gpc$^{-3}$\,yr$^{-1}$, which is in good agreement 
with the empirical estimate in  \cite{Abbott:2020khf}. 
A robust independent measurement of $R_{\rm GW190814}$ in future observing runs with GW detectors, could serve as a direct proxy for  $F_{\rm fallback}\times P_{\rm survive}$.

We now examine the possibility of the low-mass component forming first (Channel~B). 
An example binary that follows this channel is shown on the right-hand side of Fig.~\ref{fig:2} (see also Fig.~\ref{fig:a2}).  
The model is taken from the detailed binary evolution tracks of \cite{Wang:2020abc} \citep[see also][]{Langer:2019osx}, calculated at the metallicity of the Small Magellanic Cloud (SMC). 
The initial binary consists of a 44.7 and a 42.4\,M$_\odot$ star in 5-day orbit. The system first undergoes Case\,A mass transfer from the primary to the secondary, which leads to a mass-ratio inversion. The CCSN that forms the LMG object 
occurs when the separation is $\sim0.23$\,AU. The AD21 model that resembles  the pre-collapse state of the progenitor best has $M_{\rm final}=23.5\msun$ and a (higher) metallicity of $z=0.01$. This model can produce  a 2.7\msun~BH that receives a 
$\sim$400\,km\,s$^{-1}$ kick. For these parameters, the system has a  $\sim 38\%$ probability to remain bound without the LMG BH falling inside its companion at periastron, and a  $\sim$1\% chance of being kicked into a wide orbit ($\gtrsim 5$\,AU;$e\gtrsim 0.8$), which would allow the secondary to further evolve without transferring mass until the late CHeB stages. 

To better evaluate the final outcome of this evolutionary path, we used 
\mesa to  evolve a 2.7+54.9\msun~BH and MS binary model, with an 
orbit identical to the post-SN Case~B example of 
\cite{Zevin:2020gma} ($a = 9$\,AU; e = 0.95) at the SMC metallicity  
($Z=2.179\times10^{-3}$; see Appendix). The binary initiates unstable 
mass transfer and enters a CE during CHeB (at $t=9$\,Myr), 
when $a=7.7$\,AU, $e=0.94$, and  $m_2=51.9\msun$. At this stage, the 
envelope of the secondary has a binding energy of 
$E_{\rm bind}=5.8\times10^{49}$\,erg (evaluated when the Roche lobe is 
filled; $R_{\rm RL} =1047\rsun$). This means that a relatively high CE 
efficiency of $\alpha \equiv E_{\rm bind}/\Delta E_{\rm orb}\gtrsim 1.7$ 
(where $\Delta E_{\rm orb}\simeq -Gm_1 m_2^{\rm final}/2a^{\rm final}$ is the orbital energy, see \citealt{ktl+16}) is required to successfully eject the 
envelope. For $\alpha \gtrsim 2.8$, the post-CE binary remains 
detached and produces a GW190814-like 
system with a $20\msun$ BH that merges within tens of millions of years. Alternatively, if the post-CE 
separation is tighter, the binary may experience further (unstable) mass 
transfer leading to the exposure of the secondary's 15\msun~CO core. 
In this case the end-product could be either a less asymmetric BBH (with $q\simeq 
0.18$ versus $0.12$) or another LMG object or NS (in which case, the system would be disrupted).

Assuming that after the formation of the low-mass BH, the evolution of the aforementioned 
system is similar to the Channel~B example of \cite{Zevin:2020gma}, then to first order, 
the relative contribution of the two binary types to the observed merger-rate density
is determined by the frequency of the respective ZAMS progenitors. 
Assuming a \cite{Kroupa:2000iv} IMF and the \cite{Sana:2012px} distributions for the initial 
binary parameters, 
 then one would expect the merger-rate contribution of our Fig.~\ref{fig:2} example to be comparable to that in \cite{Zevin:2020gma}. This is likely conservative, given the broad mass range over 
which LMG objects form in our models.  
Nevertheless, Channel~B should be rare compared to Channel~A, mainly because it requires 
both a fine-tuned SN kick and a high CE 
efficiency. However, as already demonstrated by \cite{Zevin:2020gma}, it 
could play an increasingly important role at higher metallicities. 

Nevertheless, Channel\,A also more naturally 
explains the spin properties of GW190814 as one may expect the massive BH to have a small spin that is likely aligned with the 
orbital angular momentum, thereby resulting in a small $\chi_{\rm eff}$, consistent with what has been observed. While  the second SN may lead to misalignment, the orbit at that time is quite compact, causing the magnitude of the kick velocity to be comparable to the pre-SN orbital velocity. In turn, this should cause the  post-SN  misalignment to be small  \citep{tv22}. 
Contrarily, Channel\,B would likely produce a fast-spinning massive BH. While one would still expect the spin to be aligned with the orbital angular momentum, this might have produced some additional observable signature in the GW signal  \citep[i.e. $\chi_1 \neq 0$;  ][]{Qin:2018vaa}. 
The small $\chi_{\rm eff}$  in our scenario  would also explain why the spin properties of GW190814 resemble those of BBH mergers with $q\sim 1$, and significantly deviate from the recently reported anti-correlation between $q$ and $\chi_{\rm eff}$ found for the general BBH population  \citep[derived for mergers with $q$ in 
the $0.4-1$ range and excluding GW190814; see][]{Callister:2021abc}.

\section{Summary}\label{sec:5}
In this work we investigated the potential role of explodability islands in 
the formation of highly asymmetric-mass mergers such as GW190814 and the recently announced GW200210. 
We used a dense grid of SN progenitor models 
at different metallicities \citep[presented in detail in][]{aguilera-dena2021} to 
demonstrate that both NSs and LMG objects --- which are assumed to be BHs here --- may be 
produced from a broad range of massive stars, with \mco~ masses 
extending up to $\sim30\msun$ (Section~\ref{sec:3}; Figure~\ref{fig:1}). 
This picture agrees well with   
several recent studies \citep[e.g.][ and references above]{Laplace:2021abc,Schneider:2020vvh}, 
but it is in a stark contrast to standard BPS prescriptions, in which LMG objects may only 
form from an extremely limited range of progenitors with pre-collapse $\mco\simeq4\msun$.  
In Section~\ref{sec:4}, we demonstrate that a 
direct consequence of this modified explosion 
landscape is that asymmetric GW190814-like 
mergers may form in nearly identical way to 
`canonical' equal-mass BH mergers \citep[e.g.][]{Belczynski:2016obo}, the only 
difference being that the second core collapse event makes a LMG object instead of a 
high-mass BH. If this is indeed the case, then the merger rate of asymmetric systems that 
formed via this path should be a sizable fraction of the total BBH rate, proportional 
to the number of high-mass stars that fall inside explodability islands 
(Section~\ref{sec:4}; Eq.~\ref{eq:1}). A major difference is that, on-average, fallback 
CCSNe are expected to yield sizable natal kicks. Combining our AD21 
progenitor models with the semi-analytic core-collapse prescription of 
\cite{Mandel:2020abc}, we find that LMG objects may receive kicks ranging 
from 0 to $\sim$1200\kms. Our results for the highest 
magnitude kicks (Section~\ref{sec:4}, Figure~\ref{fig:a3}) 
suggest that $\mathcal{O}(50\%)$ of systems evolving via this channel should 
be able to survive the SN explosion and 
produce a successful merger. These results motivate further  
BPS simulations with updated final-to-remnant mass relations 
and kick distributions. Such studies should be able to further constrain the merger-rate densities of asymmetric-mass 
mergers and the (likely negligible) consequences for the LMG in Galactic binaries. 
Our Channel~B estimates suggest that this path is rare 
compared to Channel~A. Nevertheless, it may play an  
important role in high-metallicity environments and may also 
be relevant to the formation of binary compact objects in the Galaxy 
\citep[e.g. BH binaries with mildly recycled pulsar companions, or 
ultra-massive X-ray binaries such as GX\,301$-$2][]{Doroshenko:2009ei}. 
 Similarly, NSs and LMG objects originating 
 from stars with $\mco \ge 10\msun$ may 
 also contribute to mergers that are less 
 asymmetric than GW190814 and GW200210. For instance, 
 the $7 \lesssim \mco/\msun \lesssim 15$ 
 regime, which may contribute objects with 
 masses of $\sim 1.4, 3$ and $\sim 8\msun$ 
 (Figure~\ref{fig:1}), could be relevant to 
 the first two NSBH mergers GW200115 and GW200105 (component masses of 
 \citep[ $\sim $1.6+6 and $1.9+8.9\msun$,  respectively][]{Vigna-Gomez:2021oqy,Broekgaarden:2021hlu}. 
 A more detailed exploration of the entire parameter space shown in Figure~\ref{fig:1} 
 is necessary to investigate the full range of possible outcomes, 
 as well as potential contradictions to observations.  

As a final remark we note that fallback SNe leading to the 
formation of NSs or LMG objects in symmetric explosions, may also 
carry  broader implications for compact-object 
populations. For instance, they might contribute  to the 
observed pulsar population in globular clusters, thereby serving as a substitute for NSs that 
formed via electron-capture SNe, which may be rarer than previously thought  \citep{Antoniadis:2020abc}.

\begin{acknowledgements}
This work was supported by the Stavros Niarchos Foundation (SNF) and the 
Hellenic Foundation for Research and Innovation (H.F.R.I.) 
under the 2nd Call of ``Science and Society'' Action Always 
strive for excellence -- ``Theodoros Papazoglou’' (Project Number: 01431). 
A.V.-G. acknowledges support by the Danish National Research Foundation (DNRF132).
Simulations in this paper made use of the \texttt{COMPAS} 
rapid binary population synthesis code (version 02.22.00), which is 
freely available at \url{http://github.com/TeamCOMPAS/COMPAS} 
\end{acknowledgements}

\bibliographystyle{aa}
\bibliography{gw190814}

\begin{appendix}\label{supplement}
\section{Supplementary material}

\subsection*{\mesa simulations}
\begin{figure*}
\resizebox{\hsize}{!}{\includegraphics{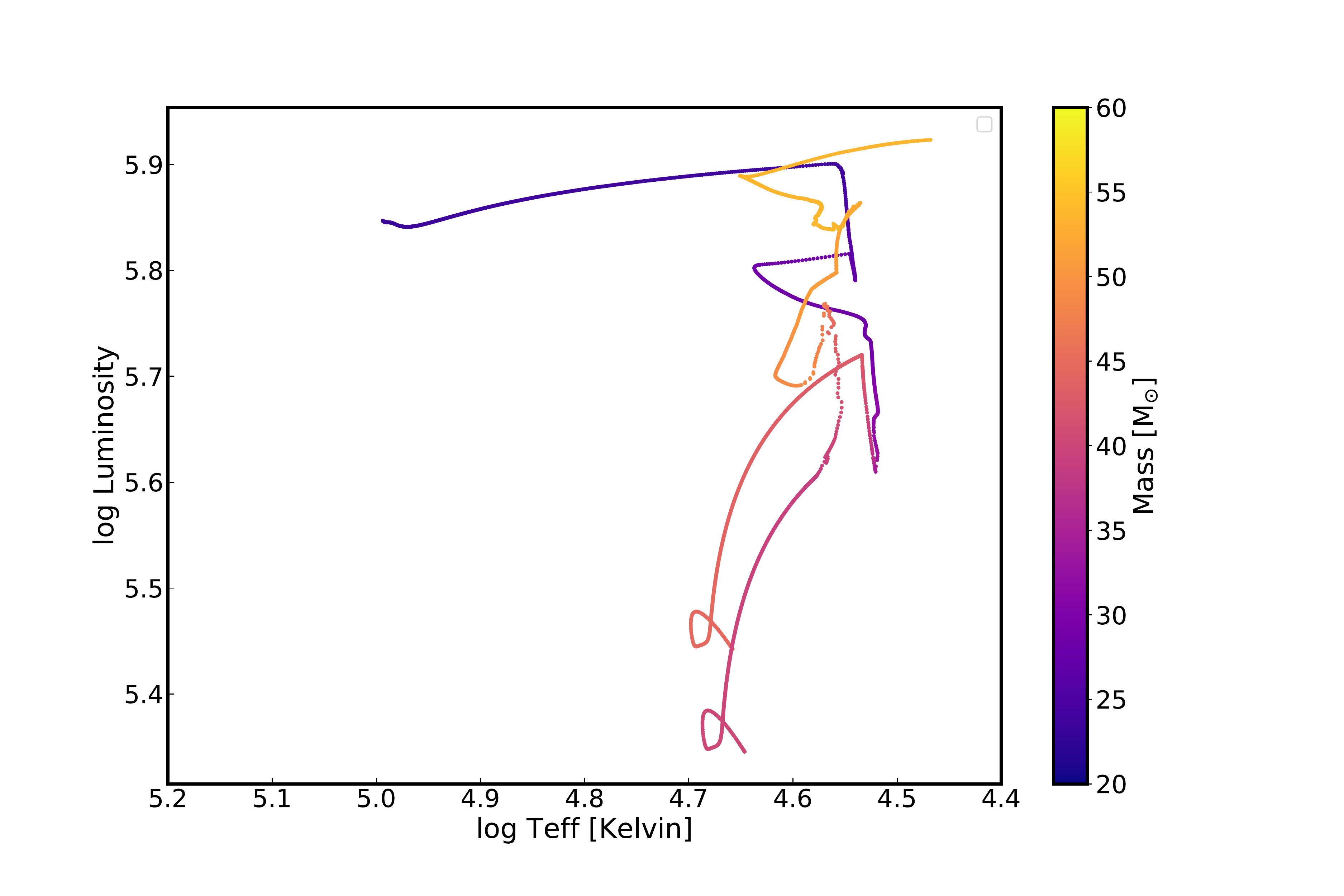}}
\caption{Evolution of our Channel B progenitor on the HR diagram from ZAMS until the formation of the first compact object (see text for details)} \label{fig:a2}
\end{figure*}

Here, we briefly describe the set-up and input physics of our 
\mesa  simulations for the Channel~B progenitor discussed in Section~\ref{sec:4}. 
The initial system is a 44.7+42.7\msun ~ZAMS pair with an SMC metallicity 
($Z=2.179\times10^{-3}$), separated by 0.15\,AU. The 
evolution of the binary up to the formation of the first compact object was 
calculated by \cite{Wang:2020abc}  \citep[see also][]{Langer:2019osx} and is 
illustrated in the HR diagram  in Figure~\ref{fig:a2}.

The subsequent evolution was modelled using \mesa~\texttt{v15140} 
\citep[see][and references therein]{Paxton:2019abc}. 
The input parameters and results for these computations are publicly 
available. Here we summarise some of the critical assumptions (where 
these deviate from the default \mesa options).
Throughout our calculations, we used the \texttt{type2} opacity 
tables of \mesa. Following \cite{Wang:2020abc}, we modelled 
convection using standard mixing-length theory, setting the mixing 
length parameter to $a_{\rm mlt} \equiv l/H_{\rm p} = 1.5$, where 
$H_{\rm p}$ is the local pressure scale height. 
Convective stability was evaluated using the Ledoux criterion, 
while we also considered the effects of convective core 
overshooting (modelled as a step function with 
$a_{\rm OV}=0.335$), thermohaline mixing ($a_{\rm th}=1.0$) 
and semi-convection ($a_{\rm sc} = 0.01$. Finally, we took into 
account rotational mixing employing the same parameters as 
\cite{Wang:2020abc}. Mass loss due to stellar winds was modelled 
using the \texttt{`Vink'} recipes \citep{Vink:2001cg,Vink:2017ujd}.

We started by evolving a 54.94\msun\ star at an SMC 
metallicity  from pre-MS to ZAMS. The ZAMS model was then placed 
in a 10\,AU, $e=0.95$ orbit around a 2.7\msun\ 
point-mass companion. Here, we adopted the default \mesa
options and parameters to simultaneously solve for the stellar 
structure and orbital evolution, taking  mass loss 
due to Roche-lobe overflow into account \citep{Paxton:2015jva}. 
The evolution of the eccentricity was inferred using the relations of \cite{Zahn:1977abc} for radiative envelopes. Once mass transfer became unstable 
($\mdot> 0.1$\msun\,yr$^{-1}$), the hydrogen-rich envelope was removed using a 
constant wind of ($\mdot= 0.1$\msun\,yr$^{-1}$). The remaining 20.37\msun\ CHeB 
helium-rich star was then left to evolve until the onset of silicon burning, 
neglecting further binary interactions. 

The evolution of the system from ZAMS until the final merger of the two compact 
objects is summarised in Figure~\ref{fig:2}. Figure~\ref{fig:a2} illustrates 
the evolution of the \cite{Wang:2020abc} model on the HR diagram 
(from ZAMS until the formation of the first compact object). 

\begin{figure*}
\centering
\begin{tabular}{cc}
\includegraphics[width=0.45\textwidth]{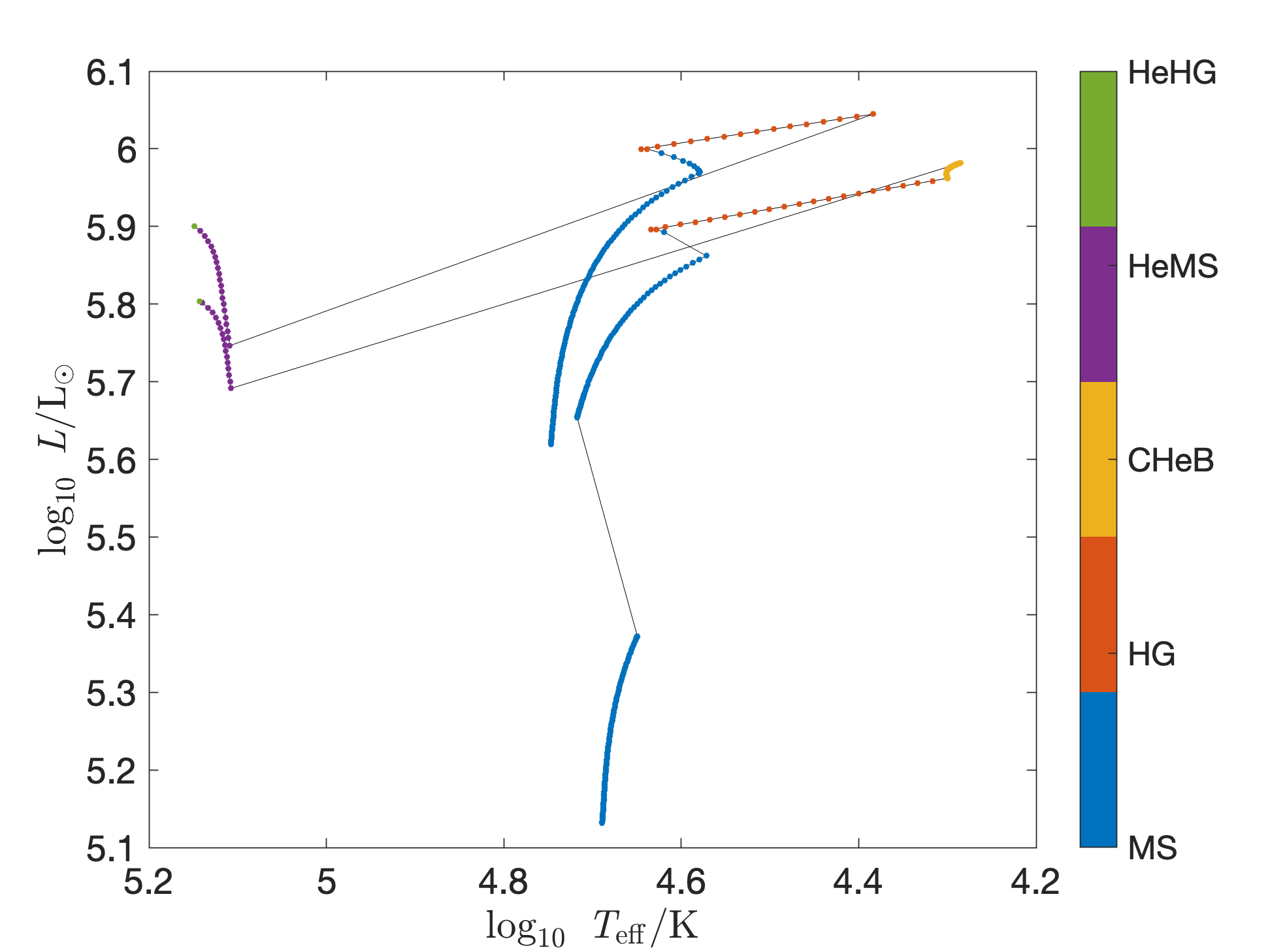} & 
\includegraphics[width=0.45\textwidth]{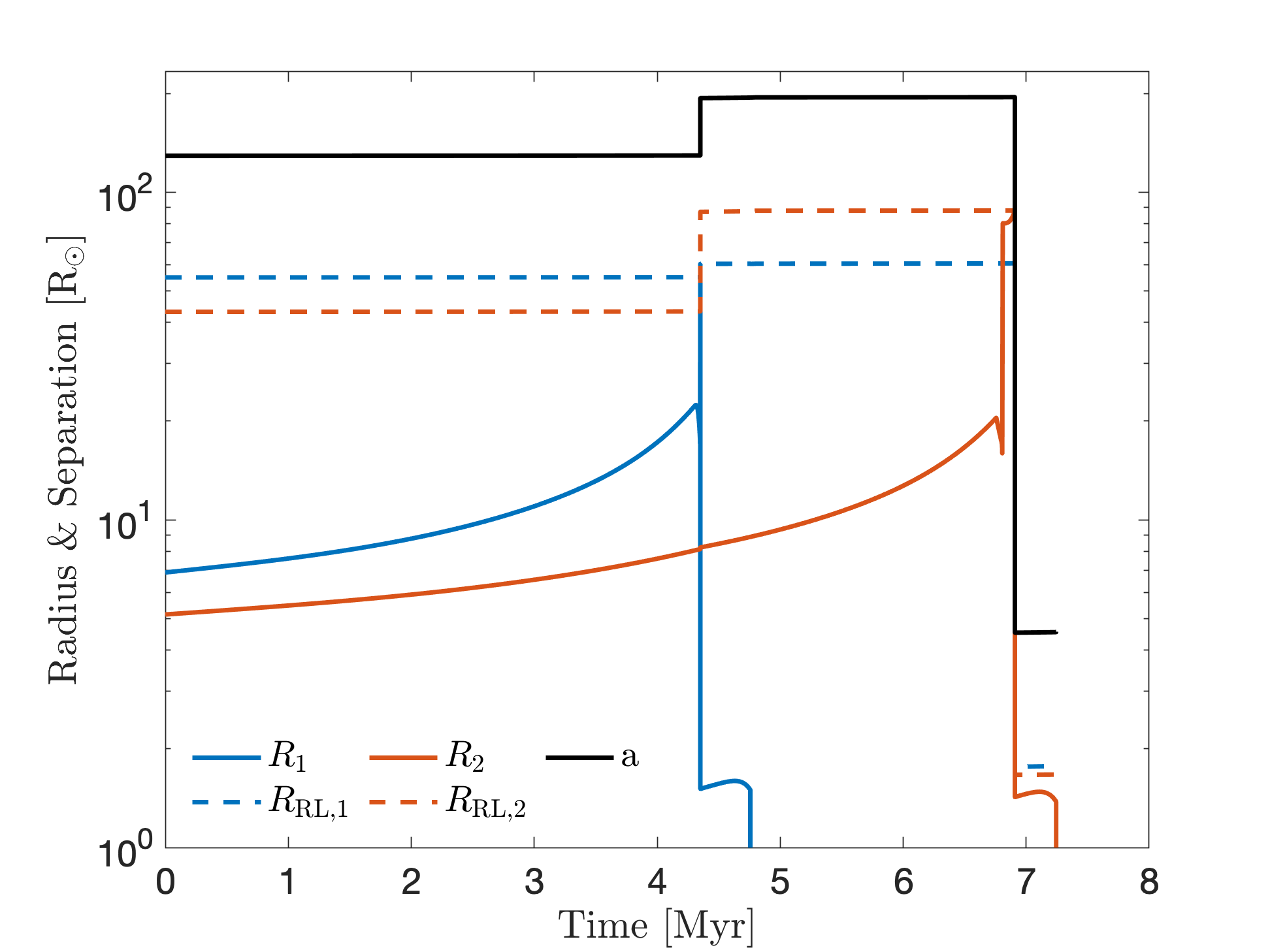} \\
\includegraphics[width=0.45\textwidth]{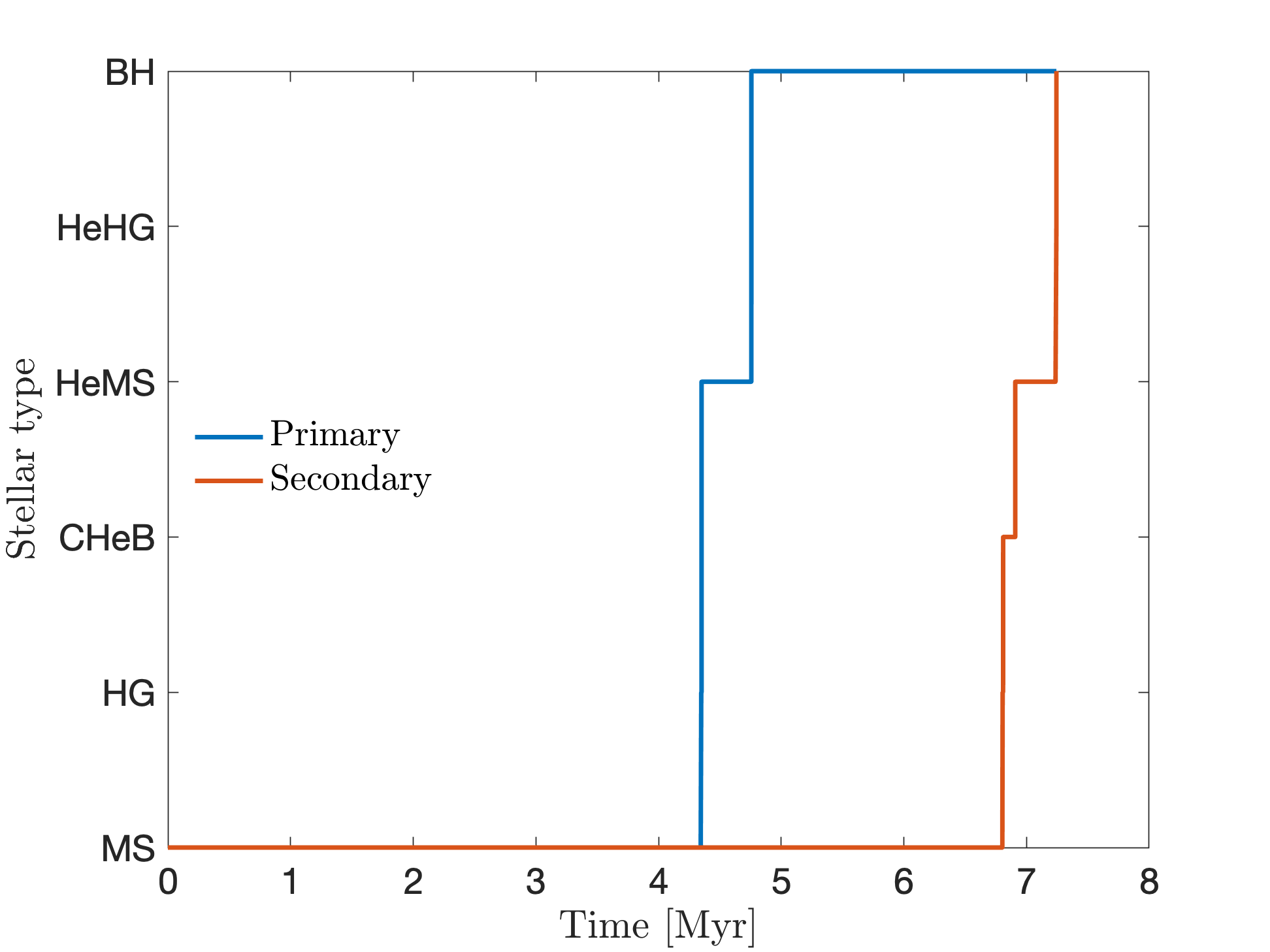} & 
\includegraphics[width=0.45\textwidth]{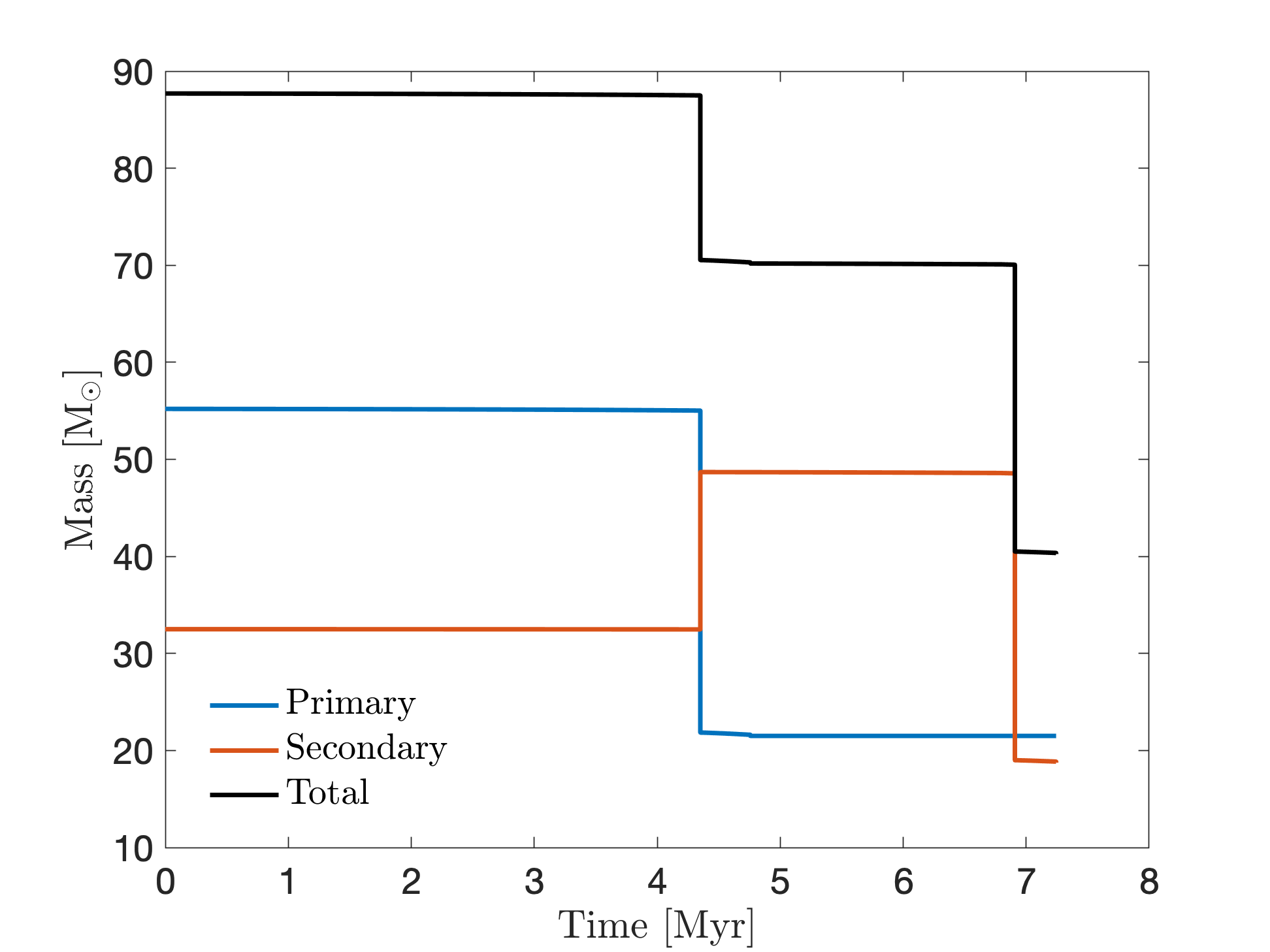} 
\end{tabular}
\caption{Summary of the evolution of the Channel~A progenitor 
system discussed in the main text. The initial binary consists of a 55.2+32.5\msun~ZAMS pair, separated by 0.6\,AU. The 
evolution of the system was calculated with the BPS code 
\texttt{COMPAS}. (Abbreviations: MS: main sequence; 
HG: Hertzsprung gap; CHeB: core helium burning; HeMS: helium 
main sequence; HeHG: Helium Hertzsprung gap; BH: black hole, $R_{i}$: radius of the $i-$th star; $R_{\rm RL,i}$: Roche-lobe radius of the $i-$th star, $a$: orbital separation).} \label{fig:a1}
\end{figure*}

\begin{figure*}
\resizebox{\hsize}{!}{\includegraphics{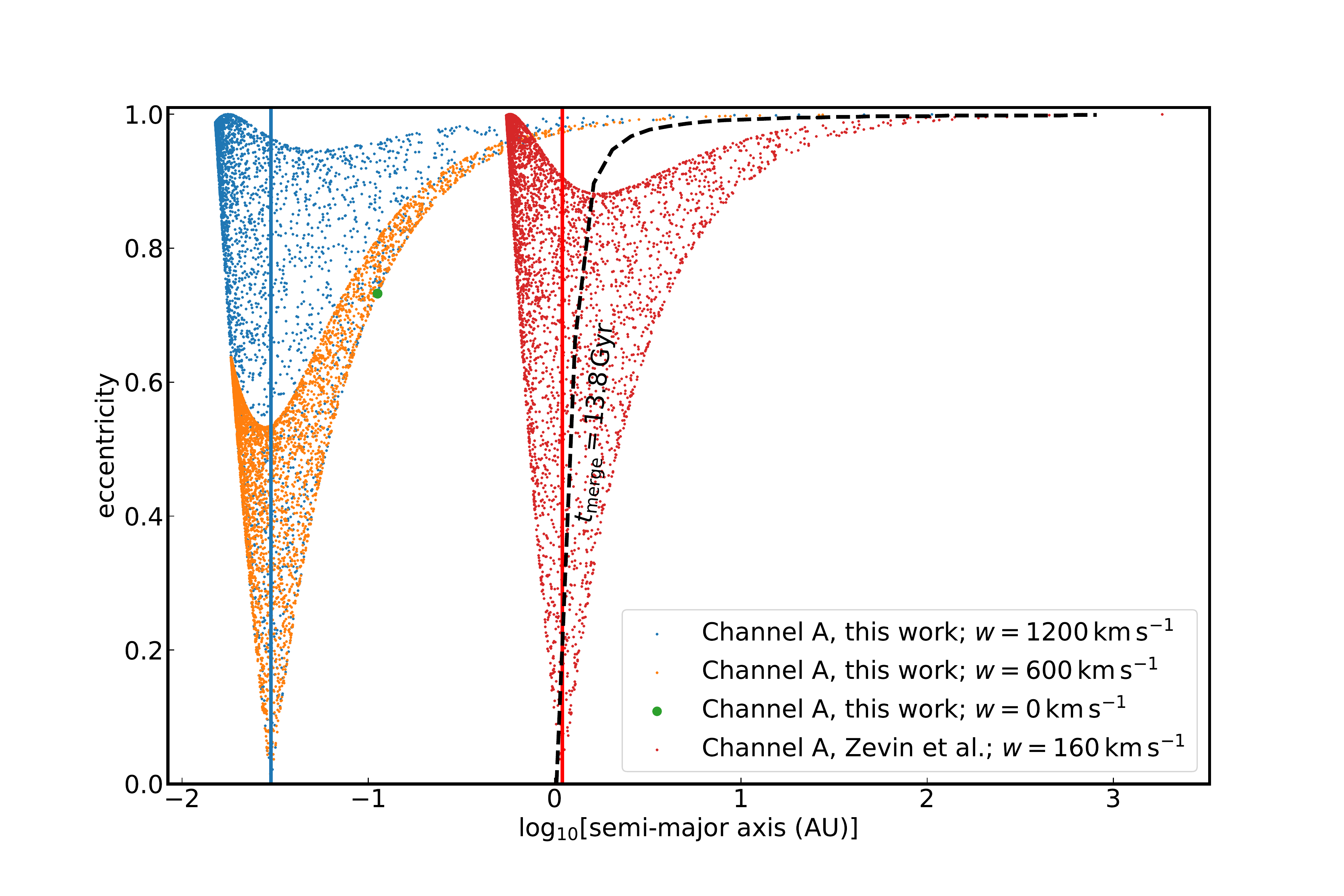}}
\caption{Influence of natal SN kicks on the post-SN binary 
configuration. The scatter plots illustrate the results of a Monte-Carlo simulation for the post-SN distribution of 
orbital configurations, assuming various initial 
configurations. Each colour represents a given pre-SN orbit, 
receiving a kick of fixed amplitude. The kick angles are 
drawn from uniform distributions \citep[see text and][]{Tauris:2017omb}. Blue, orange, and green points 
represent the Channel\,A example discussed in the main text 
when the second LMG BH is formed. Red points show 
simulations for the Channel\,A progenitor discussed in \cite{Zevin:2020gma}. Blue and red vertical lines indicate the initial orbital separations of the afforementioned systems. Binaries to the left of the 
black-dotted line merge within a Hubble time. } \label{fig:a3}
\end{figure*}

\begin{figure*}
\resizebox{\hsize}{!}{\includegraphics{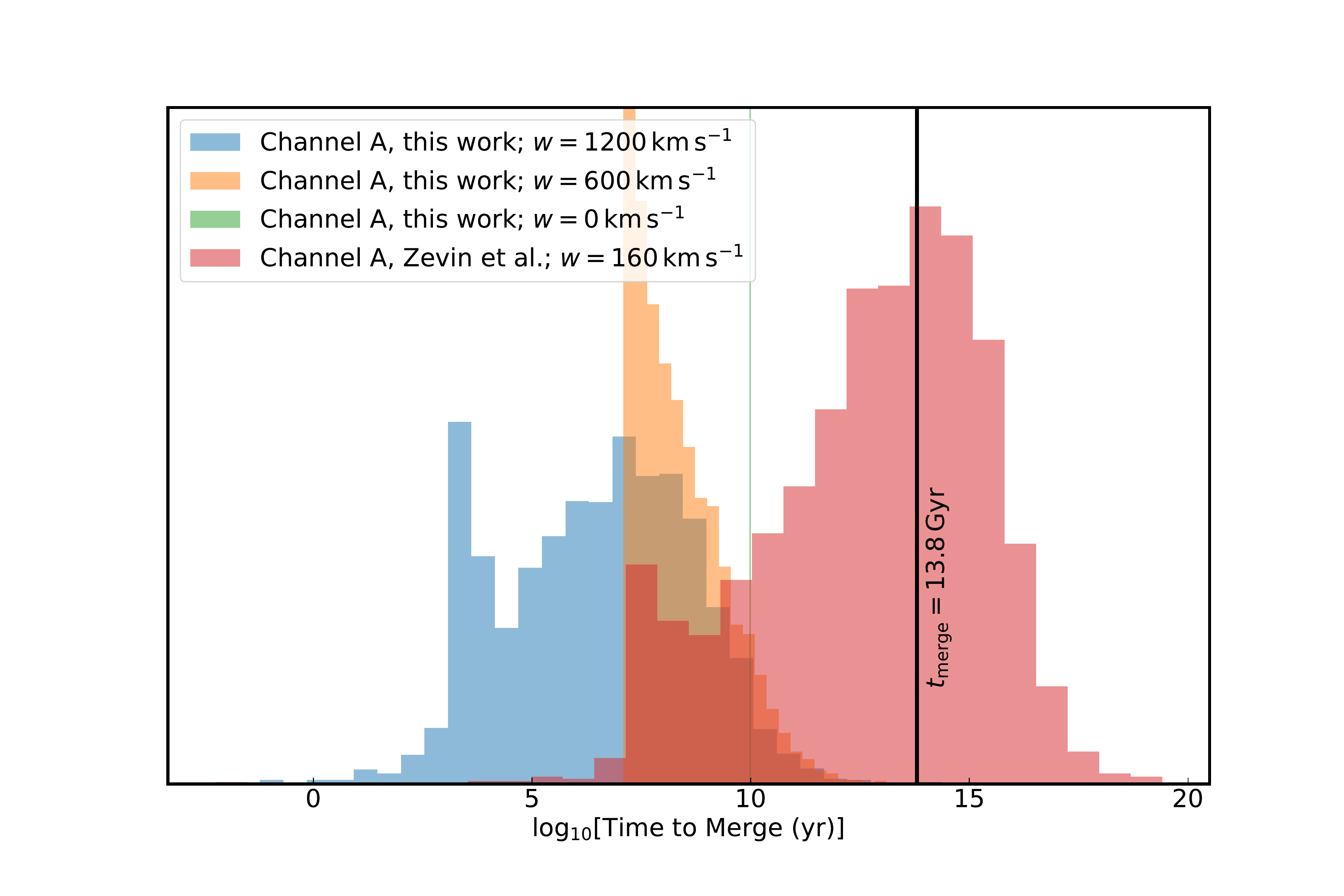}}
\caption{Distribution of merge timescales for the binary configurations shown in Fig.~\ref{fig:a3} and discussed in the main text.} \label{fig:a4}
\end{figure*}

\subsection*{Dynamical effects of SNe, and post-SN orbital evolution}
The effects of natal SN kicks on the binary orbit were calculated using the standard  prescriptions of  \cite{Hills:1983abc} and \cite{Tauris:2017omb}, which are summarised here for convenience. 

The relation between the pre- and post-SN semimajor axes is given by the following:
\begin{equation}
\frac{a_{\rm f}}{a_{\rm i}} = \left[ \frac{1- \Delta M / M}{1 - 2\Delta M/M - (w/v_{\rm rel})^2 - 2\cos\theta(w/v_{\rm rel})}\right],    
\end{equation}
where $\Delta M$ is the amount of mass ejected instantaneously ($ \ll P_{\rm orb}$) 
during the SN, $M$ is the total mass of the pre-SN system, 
$v_{\rm rel}$  is the relative velocity 
between the two stars at the time of explosion (assumed to be equal to the relative orbital velocity, $\sqrt{GM/a_{\rm i}}$, for $e=0$), $w$ is the magnitude of the kick velocity, and $\theta$ is the angle between the kick velocity vector and the pre-SN orbital plane vector. The post-SN eccentricity is given by the following:
\begin{equation}
    e = \sqrt{1 + \frac{2E_{\rm orb, f}L^2_{orb,f} }{\mu_{\rm f} G^2 M_{\rm f,1}^2 M_{\rm f,2}^2}},
\end{equation}
where $L_{\rm orb,f} = a_{\rm i} \mu_{\rm f} 
\sqrt{(v_{\rm rel} + w \cos\theta)^2 + (w\sin\theta\sin\phi)^2}$, with $\phi$ being the kick angle on the plane perpendicular to the pre-SN orbit, $\mu_{\rm f}$ is the post-SN reduced mass, and $E_{\rm orb,f}=-GM_{f,1}^2M_{f,2}^2/2a_{\rm f}$ is the post-SN orbital energy. 
The probability for the system to remain bound is as follows:
\begin{equation}
    P_{\rm bound} = \frac{1}{2}\left[ 1 + \left[ \frac{1-2\Delta M /M - (w/v_{\rm rel})^2}{2(w/v_{\rm rel})}  \right] \right].
\end{equation}
Finally, a given post-SN system is assumed to survive only if the semi-minor axis is larger than the Roche-lobe radius of the secondary,
\begin{equation}
    \lambda \equiv a(1-e^2) \leq R_{\rm RL,2}. 
\end{equation}

Figure~\ref{fig:a3} illustrates simulated post-SN systems using the equations above and assuming a kick of a fixed magnitude applied in a random uniform direction. The black-dashed line shows the combination of orbital parameters for which the time to merge equals the Hubble time ($13.8$\,Gyr). The latter was evalulated numerically following \cite{Peters:1964zz} \citep[see also][]{Tauris:2017omb}:

\begin{eqnarray}
    \tau_{\rm merge}(a_0,e_0) = \frac{12}{19}\frac{C_0^4}{\beta} \nonumber &\\
    &\times \int_0^{e_0}{\frac{e^{29/19}[1 + (121/304)e^2]^{1181/2299}}{(1-e^2)^{3/2}}de},
\end{eqnarray}
where 
\begin{equation}
    C_0 = \frac{a^0(1-e_0^2)}{e^(12/19)_0}[1 + (121/304)e_0^2]^{-870/2299}
\end{equation}
and
\begin{equation}
    \beta = \frac{64G^3}{5c^5}M^2\mu. 
\end{equation}
Figure~\ref{fig:a4} shows the corresponding $\tau_{\rm merge}$ distributions for the aforementioned simulations.

\end{appendix}

\end{document}